\documentclass[twocolumn,aps,prxquantum,superscriptaddress,longbibliography]{revtex4-2}
\usepackage[normalem]{ulem}
\usepackage{graphicx}
\usepackage{amsmath}
\usepackage{amssymb}
\usepackage{braket}
\usepackage{physics}
\usepackage{ulem}
\usepackage{bm}
\usepackage{dcolumn}
\usepackage{color}
\usepackage[colorlinks=true,linkcolor=blue,citecolor=blue,urlcolor=blue]{hyperref}

\begin{document}

\title{Emergent Gauge Theory in Rydberg Atom Arrays}
\author{Yanting Cheng}
\affiliation{Institute of Theoretical Physics and Department of Physics, University of Science and Technology Beijing, Beijing 100083, China}

\author{Hui Zhai}
\email{hzhai@tsinghua.edu.cn}
\affiliation{Institute for Advanced Study, Tsinghua University, Beijing 100084, China}
\affiliation{Hefei National Laboratory, Hefei 230088, China}
\date{\today}

\begin{abstract}

Rydberg atom arrays have emerged as a novel platform exhibiting rich quantum many-body physics and offering promise for universal quantum computation. The Rydberg blockade effect plays an essential role in establishing many-body correlations in this system. In this review, we will highlight that the lattice gauge theory is an efficient description of the Rydberg blockade effect and overview recent exciting developments in this system from equilibrium phases to quantum dynamics. These developments include realizing exotic ground states such as spin liquids, discovering quantum many-body scar states violating quantum thermalization, and observing confinement-deconfinement transition through quantum dynamics. We emphasize that the gauge theory description offers a universal theoretical framework to capture all these phenomena. This perspective of Rydberg atom arrays will inspire further the future development of quantum simulation and quantum computation in this platform.   
    
\end{abstract}

\maketitle

Gauge symmetry is a fundamental symmetry in nature, which plays an essential role in the standard model unifying electromagnetic, weak, and strong forces \cite{Polyakov1987}. Gauge symmetry also appears as an emergent symmetry in the low-energy effective theory of strongly correlated condensed matter systems. The most well-studied examples include fractional quantum Hall effect \cite{Shouchenprl,review_of_quantum_Hall}, cuprate superconductors \cite{review_of_cuprate_superconductor}, and spin liquids \cite{review_spin_liquid_leon,wen2007,Zhou_review_2017}. The low-energy physics of these strongly correlated phases can be well captured by studying the gauge structure of these theories.

Rydberg atoms have recently emerged as a new platform for modern quantum science and technology \cite{Saffman2010,Saffman2016,Browaeys_quantum2020,Browaeys_NP2020,Morgado2021,You2021}. Rydberg atom is an excited state with a very high electron principle quantum number $n$. Using alkali atoms as an example, the radius of Rydberg atoms and the transition dipole moment between adjacent Rydberg states both scale as $n^2$, resulting in the van der Waals interaction between two Rydberg states scaling as $n^{11}$ \cite{Browaeys_NP2020,You2021}. This leads to a significantly strong dipolar interaction between two Rydberg states. Moreover, the radiative lifetime of the Rydberg excited state scales as $n^3$, which leads to a metastable state with a lifetime of the order of $100\mu s$, four orders of magnitude longer than low-lying excited states \cite{Browaeys_NP2020,You2021}. In addition,  optical tweezers offer the flexibility in arranging and moving these Rydberg atoms into arrays with different geometries \cite{Browaeys_quantum2020}. Because of these advantages of strong dipole interaction, relatively long lifetime, and flexibility in spatial control, Rydberg atom arrays have now played an increasingly significant role in studying quantum many-body physics and realizing universal quantum computation \cite{Beugnon2007,Kim2016,Browaeys2016nature,Browaeys2016science,Lukin2016science,Lukin2017nature,Lienhard2018prx,Lukin2018prl,lukin2019science_cat,Lukin2019nature,Lukin2019prl2,Browaeys2020pra,Endres2020np,scar-Floquet,Browaeys2021nature,Lukin2021nature,experiment_spin_liquid,Lukin2022nature,Dan2022PRL,Endres2023nature,Norman2023squeezing,Ye2023squeezing,Bernien2023science,Graham2023,Lukin2023nature,Scholl2023,Ma2023,Lukin2023nature2,Joanna2023,Bloom2023}.

Recent experiments have discovered both intriguing low-energy equilibrium states and novel non-equilibrium dynamics in this platform. These phenomena include different quantum phase transitions and the Kibble-Zurek scaling across the transitions \cite{Browaeys2016nature,Lukin2017nature,Lienhard2018prx,Lukin2019nature,Lukin2021nature,Browaeys2021nature}, quantum spin liquid states \cite{experiment_spin_liquid} and quantum many-body scar states \cite{Lukin2017nature,scar-Floquet}. The strong correlations behind these phenomena originate from the Rydberg blockade effect due to strong dipole interaction between Rydberg atoms. This review aims to cover these interesting experimental observations due to Rydberg blockade effects in atom arrays and to provide a unified theoretical view of these observations. The theoretical view turns out to be an emergent gauge theory, which naturally captures the Rydberg blockade-induced correlation effects both in- and out-of-equilibrium \cite{PXP-gauge,Cheng-PRXQ,pan,Cheng-NJP,Halimeh2023quantum,Halimeh2023prb}. This theoretical view, on the one hand, highlights that the emergent gauge theory is a common feature in strongly correlated many-body systems, such as cuprates and fractional quantum Hall effect. On the other hand, it brings out the connection between the Rydberg atom arrays and recent developments in quantum simulations of gauge fields (for a review, see \cite{Jad2023review}). As we will discuss in detail later, in the simplest situation of Rydberg atom array, the emergent gauge theory is the same as the one-dimensional lattice Schwinger model realized in recent cold atom quantum simulator \cite{USTC,Yuan2022science,USTC-thermalization}, from which we can gain the physical intuition and understanding to help us with more complicated geometries. 

\textit{Rydberg Blockade Effect and the PXP Model.} In this review, we focus on the situation that optical tweezers completely fix the spatial positions of atoms, and we only need to consider the internal degree of freedom of each atom. Here we focus on the situation that each tweezer only traps one single atom, therefore, the ground state and a Rydberg excited state form a pseudo-spin doublet, with $\ket{\uparrow}$ being the Rydberg state and $\ket{\downarrow}$ being the ground state. The optical tweezers can arrange atoms into any geometry in space. For simplicity, we first discuss the situation that all tweezers are lined up into a one-dimensional array. The lessons we learned from this simple geometry can be generalized to more complicated geometry later. The Hamiltonian for a one-dimensional Rydberg atom array can be written as  
 \begin{eqnarray}          
 \hat{H}=\Omega\sum_i\hat{S}^x_i-\Delta\sum_i\hat{S}^z_i+\sum_{i<j}V_{ij}\hat{n}_i\hat{n}_j. \label{H1}
\end{eqnarray}
The $\hat{S}^x$ term denotes the coupling between the ground state and the Rydberg state, usually provided by a two-photon process with coupling strength $\Omega$. The $\hat{S}^z$ term denotes the energy difference between the Rydberg excitation energy and the two-photon energy, with $\Delta$ being the detuning. 

The Rydberg atom number $\hat{n}_i$ at each site is given by $\hat{n}_i=\hat{S}^z_i+1/2$. The last term in Eq. \ref{H1} only includes the van der Waals interaction between Rydberg atoms, and 
\begin{equation}
V_{ij}=\frac{C_6}{|r_i-r_j|^6}.
\end{equation}
We introduce a length scale called the Rydberg blockade radius $R_\text{b}=(C_6/\Omega)^{1/6}$, which typically is a few microns. When $|r_i-r_j|< R_\text{b}$, the van der Walls interaction between two Rydberg atoms can be much larger than $\Omega$ and $|\Delta|$. Hence, it prevents two Rydberg atoms from sitting within the blockade radius, an effect well known as the Rydberg blockade. Here we first consider a situation in which $a$ is smaller than $R_\text{b}$ but $2a$ is larger than $R_\text{b}$. Since the Rydberg interaction decays as $1/r^6$, we can have a situation in which $C_6/a^6$ is larger than $\Omega$ and $|\Delta|$ by about an order of magnitude, while $C_6/(2a)^6$ is smaller than $\Omega$ and $|\Delta|$ by about an order of magnitude. Hence, we can make a bold approximation for studying the low-energy physics by simplifying $V_{ij}$ as  
 \begin{displaymath}
V_{ij} =\left\{ \begin{array}{ll}
 \infty, & \text{when} \   \  i=j\pm 1, \\
0, & \text{when} \  \ i\neq j\pm 1.
 \end{array}\right.
\end{displaymath}
Moreover, we note that the interactions between Rydberg and ground-state atoms, as well as between two ground-state atoms, are completely negligible when two atoms are separated by several microns in space because these interactions are smaller than the interaction between two Rydberg atoms by many orders of magnitudes\cite{Browaeys_NP2020,You2021}.  

This simplification captures the essence of the Rydberg blockade effect when $a<R_\text{b}<2a$, which prevents two Rydberg atoms from sitting at neighboring sites in the low-energy Hilbert space. This constraint can be implemented by applying a projection operator $\hat{\mathcal{M}}$ to the Hamiltonian Eq. \ref{H1}, and the projection operator $\hat{\mathcal{M}}$ reads
\begin{equation}
\hat{\mathcal{M}}=\prod\limits_{i}(1-\hat{n}_i\hat{n}_{i+1}).
\end{equation}
Applying this projection operator to the Hilbert space, the low-energy effective Hamiltonian on the restricted Hilbert space reads 
\begin{eqnarray}
 \hat{H}_{\textrm{PXP}}=\Omega\sum_i\hat{P}_{i-1}\hat{S}^x_i\hat{P}_{i+1}-\Delta\sum_i\hat{S}^z_i.
 \label{H_PXP}
\end{eqnarray}
Note that $1-\hat{n}_i$ is a projection operator to project out the Rydberg state at site-$i$, and we denote this operator as $\hat{P}_i$. This Hamiltonian is now well-known as the PXP model in recent literature \cite{Subir2004prb,PXP_pra2012,Papic2018np,Papic2018prb,papic2018prb1,PXP_condensation2019,Lukin2019prl,PXP1,Papic2020prb,Papic2021np,Papic2021prx,Bernevig2022}.

\textit{Emergent Gauge Theory.}  Before we discuss the emergent gauge theory in the PXP model, we first remind the readers how gauge theory emerged in the Fermi-Hubbard model when this model was broadly discussed in the context of cuprate superconductivity. 

The Hubbard model considers spin-$1/2$ fermions in lattice. When the Fermi Hubbard model is used to describe the cuprate superconductor, the on-site repulsion between two fermions with opposite spins is the largest energy scale in the problem, which forbids double occupation of any site. In other words, at each site, the dimension of the low-energy Hilbert space equals three: empty or singly occupied by a fermion with spin-$\sigma$ ($\sigma=\uparrow,\downarrow$). This constraint can be formulated as an inequality $\hat{n}_{i\uparrow}+\hat{n}_{i\downarrow} \leq1$. 

Usually, it is not convenient to deal with these inequality constraints. To this end, slave particle methods are invented to turn these inequality constraints into equality constraints. To be concrete, we can introduce spinless holon operator $\hat{h}_i$ that only carries charge and spinon operators $\hat{f}_{i\sigma}$ that only carry spin by writing the fermion operator as $\hat{c}_{i\sigma}=\hat{h}_i^\dagger\hat{f}_{i\sigma}$ \cite{Barnes_1976,Read_1983,Coleman1984prb,review_of_cuprate_superconductor}. Therefore, when a site-$i$ is empty, we can say that this site is occupied by a holon with $\hat{h}^\dag_i\hat{h}_i=1$, and when a site-$i$ is occupied by a fermion with spin-$\sigma$, we can say that this site is occupied by a spinor with $\hat{f}^\dag_{i\sigma}\hat{f}_{i\sigma}=1$ \cite{Barnes_1976,Read_1983,Coleman1984prb,review_of_cuprate_superconductor}. Hence, the inequality constraint becomes 
\begin{equation}
\hat{h}^\dag_i\hat{h}_i+\hat{f}^\dag_{i\uparrow}\hat{f}_{i\uparrow}+\hat{f}^\dag_{i\downarrow}\hat{f}_{i\downarrow}=1. \label{constraint-Hubbard}
\end{equation}
It seems that we have enlarged the Hilbert space by decomposing a fermion into a holon and a spinon. However, with these constraints, the Hilbert space dimension at each site still equals three. 

 \begin{table*}[t]
        \centering
        \begin{tabular}{|c|c|c|}
        \hline
        &Rydberg Blockade & Hubbard Repulsion\\
        \hline
       Inequality Constraint &
        $\hat{n}_i+\hat{n}_{i+1}\leq 1$&$\hat{n}_{i\uparrow}+\hat{n}_{i\downarrow}\leq1$\\
        \hline
Enlarge       & Auxiliary Fermion Method &Slave Particle Method \\
Hilbert Space    &    $\hat{S}_{i}^-\rightarrow\hat{S}_{i}^-\hat{f}^\dagger_{i-1,i}\hat{f}^\dagger_{i,i+1}$&$\hat{c}_{i\uparrow}=\hat{h}_i^\dagger\hat{f}_{i\uparrow}$\\   
    & $\hat{S}_{i}^+\rightarrow\hat{S}_{i}^+\hat{f}_{i-1,i}\hat{f}_{i,i+1}$&$\hat{c}_{i\downarrow}=\hat{h}_i^\dagger\hat{f}_{i\downarrow}$\\ 
        \hline
        Equality Constraint&
        $\hat{S}^z_{i}+\hat{S}^z_{i+1}+\hat{f}^\dag_{i,i+1}\hat{f}_{i,i+1}=0$&$\hat{h}_i^\dagger\hat{h}_i+\hat{f}_{i\uparrow}^\dagger\hat{f}_{i\uparrow}+\hat{f}_{i\downarrow}^\dagger\hat{f}_{i\downarrow}=1$\\
        \hline
        Emergent 
        & $\hat{S}^+_{i}\rightarrow e^{-i\phi_i}\hat{S}^+_{i}$
         &$\hat{h}_i\rightarrow e^{i\phi_i}\hat{h}_i$\\
    Local     & $\hat{S}^+_{i+1}\rightarrow e^{-i\phi_{i+1}}\hat{S}^+_{i+1}$ &$\hat{f}_{i\uparrow}\rightarrow e^{-i\phi_i}\hat{f}_{i\uparrow}$\\
   Gauge Symmetry    & $\hat{f}_{i,i+1}\rightarrow e^{i(\phi_i+\phi_{i+1})}\hat{f}_{i,i+1}$ &$\hat{f}_{i\downarrow}\rightarrow e^{-i\phi_i}\hat{f}_{i\downarrow}$\\
        \hline
        \end{tabular}
        \caption{Comparison of emergent local gauge symmetry between the Rydberg blockade effect and the Hubbard repulsion effect. In both cases, a local inequality constraint can be turned into a local equality constraint by introducing an auxiliary degree of freedom to enlarge the Hilbert space, and the resulting local equality constraints can be viewed as a set of locally conserved quantities as manifestations of local gauge symmetries.}
        \label{tab}
        \label{gauge-theory}
    \end{table*}

These equality constraints Eq. \ref{constraint-Hubbard} can be viewed as locally conserved quantities at each site. Usually, the conserved quantity is a manifestation of certain symmetry. Therefore, it is natural to ask what kinds of symmetries give rise to these conserved quantities, and the answer turns out to be the local gauge symmetry. Since only fermion operators are physical operators that appear in the Hubbard model, it is easy to see that by making a local gauge transformation $\hat{h}_i\rightarrow e^{i\phi_i}\hat{h}_i$ and $\hat{f}_{i\sigma}\rightarrow e^{-i\phi_i}\hat{f}_{i\sigma}$, all fermion operators are invariant and therefore the physical Hamiltonian is invariant. This local $U(1)$ gauge symmetry defined at each site gives rise to the particle number conservation Eq. \ref{constraint-Hubbard} also defined at each site. 

Hence, from the discussion above, we learn why local gauge symmetry can emerge in the low-energy effective theory of a strongly interacting system. The key ingredients can be summarized into the following points: 
\begin{enumerate}
  \item The energetic considerations impose local constraints on the low-energy Hilbert space.
  \item These local constraints, usually written as inequalities, can be turned into equalities by introducing an auxiliary degree of freedom and virtually enlarging the Hilbert space.
  \item These equality constraints can be viewed as local conserved quantities as a manifestation of local symmetries.
  \item The virtually introduced auxiliary degree of freedom causes redundancy and leads to local gauge symmetries, precisely being the local symmetries in need. 
\end{enumerate}
Below, we will show that the emergence of gauge theory in the Rydberg atom arrays follows precisely the same reason.  

\begin{figure}[t]
    \centering
    \includegraphics[width=0.45\textwidth]{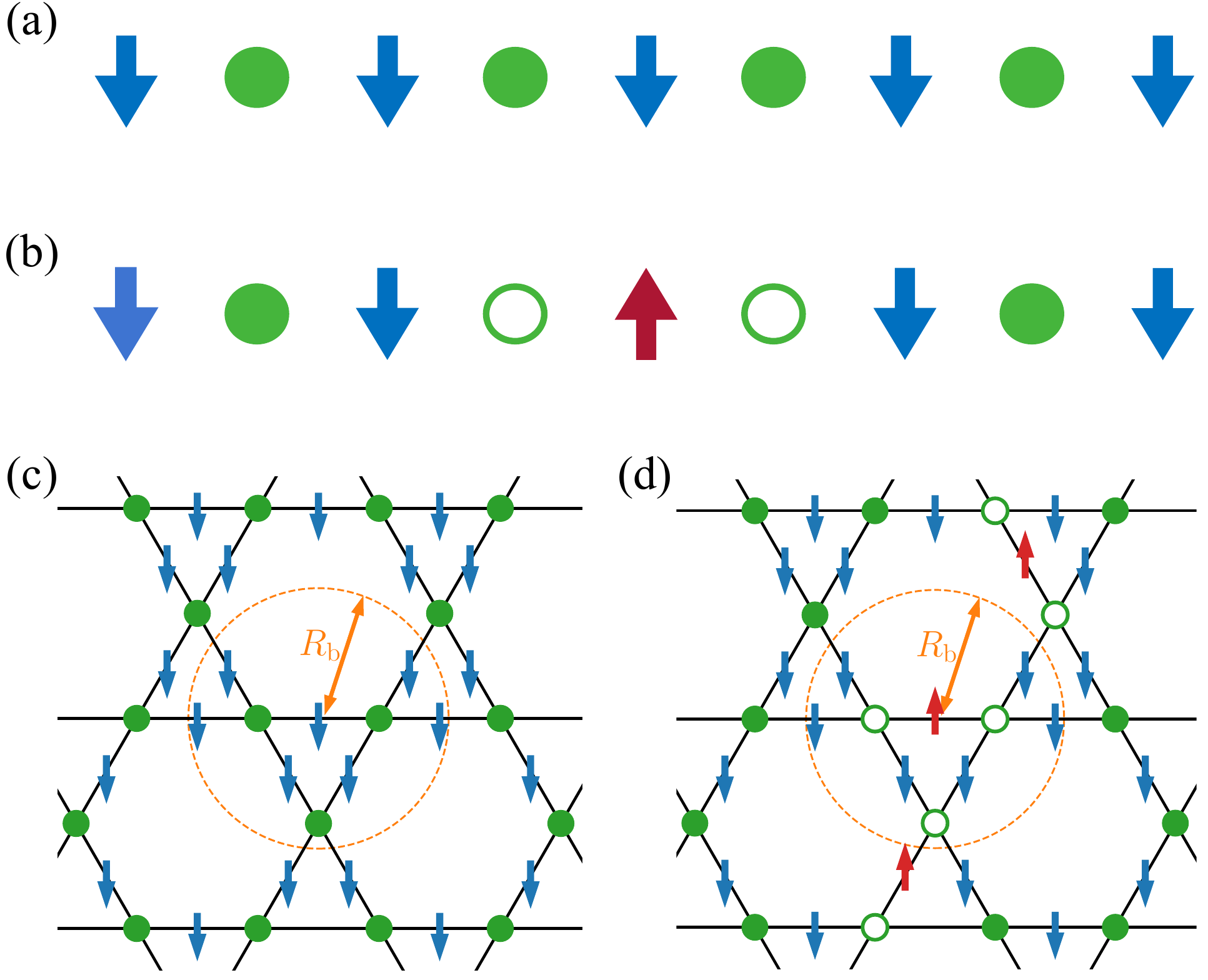}
    \caption{\textbf{Gauge theory description of Rydberg blockade effect.} (a-b) Gauge theory for one-dimensional PXP model. Auxiliary fermions are introduced on the links between two spins. (c-d) Gauge theory for Kagome lattice model. Atoms are placed on the links of the Kagome lattice. Auxiliary fermions are introduced on the sites of the Kagome lattice. The dashed circles in (c) and (d) denote the Rydberg blockade radius. In all figures, downarrows denote ground state atoms and uparrows denote the Rydberg state atoms. Open circles denote unoccupied sites and solid circles denote occupied sites. 
    \label{gauge}}
\end{figure}

The Rydberg blockade effect introduces constraints on the low-energy Hilbert space as $\hat{n}_i+\hat{n}_{i+1}\leqslant 1$ for each site. Here $\hat{n}_i$ denotes the Rydberg atom number at site-$i$. Now we introduce auxiliary fermions sitting at the links between each two sites, as shown in Fig. \ref{gauge}(a). We now rewrite the PXP term in the effective Hamiltonian as  \cite{PXP-gauge,Cheng-PRXQ,pan,Cheng-NJP,Halimeh2023quantum,Halimeh2023prb}
\begin{equation}
\hat{H}_\text{gauge}=\frac{\Omega}{2}\sum\limits_{i}(\hat{S}^{+}_i\hat{f}_{i-1,i}\hat{f}_{i,i+1}+\text{h.c.})-\Delta\sum_i\hat{S}^z_i, \label{H-gauge}
\end{equation} 
where $\hat{f}_{i-1,i}$ denotes the auxiliary fermion operator in the link between site-$(i-1)$ and site-$i$. The Hamiltonian Eq. \ref{H-gauge} says that when $\hat{S}^{+}_i$ operator flips an atom at site-$i$ from the ground state $\ket{\downarrow}$ to the Rydberg excited state $\ket{\uparrow}$, simultaneously two fermions at its neighboring links are annihilated, as shown in Fig. \ref{gauge}(b). Therefore, the Rydberg excitation at site-$(i+1)$ and site-$(i-1)$ are blocked because they share a link with site-$i$ where the auxiliary fermion is already annihilated. In other words, we utilize the auxiliary fermions to implement the Rydberg blockade effect \cite{Cheng-NJP,pan}. When atoms in two neighboring sites are both in the ground states, we have $\hat{S}^z_i+\hat{S}^z_{i+1}=-1$ and $\hat{n}^\text{f}_{i,i+1}=1$, where $\hat{n}^\text{f}_{i,i+1}=\hat{f}^\dag_{i,i+1}\hat{f}_{i,i+1}$ denotes the fermion number at the link. When one of these two atoms is excited to the Rydberg state, we have  $\hat{S}^z_i+\hat{S}^z_{i+1}=0$ and $\hat{n}^\text{f}_{i,i+1}=0$. Hence, the constraint becomes an equality as 
\begin{equation}
\hat{S}^z_i+\hat{S}^z_{i+1}+n^\text{f}_{i,i+1}=0. \label{conserved-quantity}
\end{equation} 
This constraint Eq. \ref{conserved-quantity} can be viewed as a conservation law due to a locally defined symmetry. We note that under a local gauge transformation $\hat{S}^+_{i}\rightarrow e^{-i\phi_i}\hat{S}^+_{i}$, $\hat{S}^+_{i+1}\rightarrow e^{-i\phi_{i+1}}\hat{S}^+_{i+1}$ and simultaneously $\hat{f}_{i,i+1}\rightarrow e^{i(\phi_i+\phi_{i+1})}\hat{f}_{i,i+1}$, the Hamiltonian Eq. \ref{H-gauge} is invariant \cite{Cheng-NJP,Cheng-PRXQ,pan,PXP-gauge}. It is precisely this local gauge symmetry that gives rise to the conserved quantity Eq. \ref{conserved-quantity}. As a $U(1)$ lattice gauge theory, Eq. \ref{H-gauge} can be written as the lattice Schwinger model, where the spins and fermions play the role of gauge field and matter field, respectively \cite{Schwinger_1,Schwinger_2,Coleman_AoP,Kogut_1979,Wiese1997,Kogut_1983,David,Peter2005prl,Hauke2019prl,Huang2019prl}. As a separate development, the lattice Schwinger model has also been implemented recently by using bosons in optical lattices \cite{USTC,Yuan2022science,USTC-thermalization}. 

In Table \ref{gauge-theory} we compare the emergence of gauge theories from Hubbard repulsion and the Rydberg blockade. We emphasize that these two seemingly different cases share a common mechanism. The gauge theory discussed here for the PXP model can be straightforwardly extended to one-dimensional Rydberg arrays with larger blockade radius \cite{Cheng-Li} and to the two-dimensional Rydberg arrays, including classical spin order in square and triangular lattices and the spin liquid in Kagome lattice \cite{Cheng-NJP}.

\begin{figure}[t]
    \centering
    \includegraphics[width=0.48\textwidth]{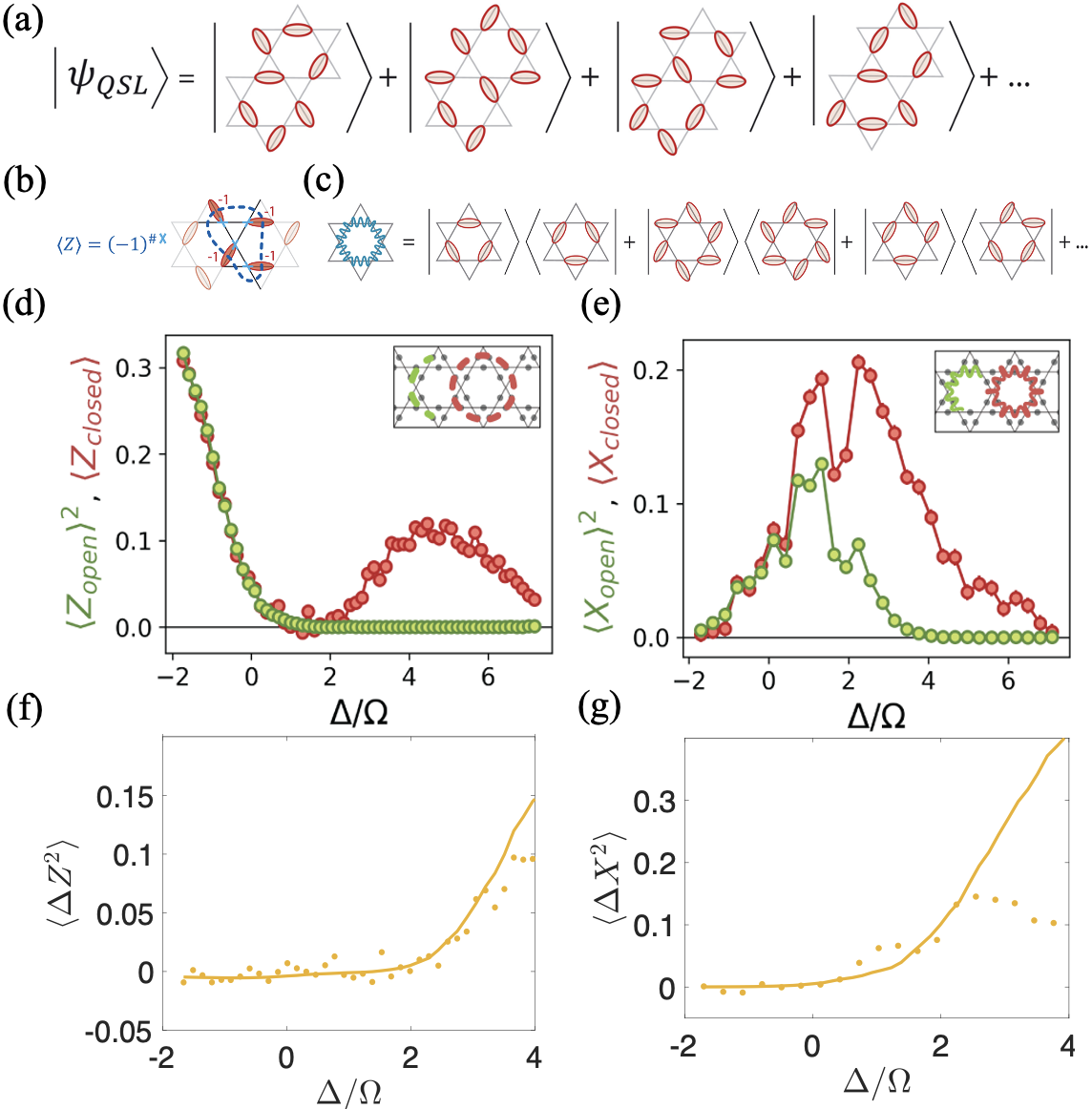}
    \caption{\textbf{Spin liquid state with Rydberg atoms array.} (a) Schematic of the spin liquid state in the system as shown in Fig. \ref{gauge}(c-d). Atoms in the bonds with red circles are excited to the Rydberg states. The spin liquid is a superposition of all closely packed configurations allowed by the blockade conditions. (b-c) $\hat{Z}$ and $\hat{X}$ operators along a closed look. $\hat{Z}$ operator shown in (b) measures the parity of Rydberg atoms along the loop. $\hat{X}$ operator couples different allowed configurations along the loop. (d-e) Experimental measurements of $\langle\hat{Z}\rangle$ (d) and $\langle\hat{X}\rangle$ (e) along a closed loop ($\langle\hat{Z}_\text{closed}\rangle$ and $\langle\hat{X}_\text{closed}\rangle$) and the results are compared with measurements along half open loop ($\langle\hat{Z}_\text{opem}\rangle^2$ and $\langle\hat{X}_\text{open}\rangle^2$). (f-g) Comparison between the experimental measurements of $\langle \Delta\hat{Z}^2\rangle  \equiv \langle\hat{Z}_\text{closed}\rangle-\langle\hat{Z}_\text{opem}\rangle^2$ (f) and $\langle \Delta\hat{X}^2\rangle  \equiv \langle\hat{X}_\text{closed}\rangle-\langle\hat{X}_\text{opem}\rangle^2$ (g) and the prediction of our variational wave function (solid line) without fitting parameters. (a-e) are reprinted from Ref. \cite{experiment_spin_liquid} and (f-g) are reprinted from Ref. \cite{Cheng-NJP}.
    \label{spin-liquid}}
\end{figure}

\textit{Ground State and Spin Liquid.} The gauge theory model Eq. \ref{H-gauge} shows that each spin flip operator is always combined with two fermion annihilation operators. In other words, each spin-$\downarrow$ is bound with two fermions at its neighboring links. This observation motivates us to write down the following wave function ansatz as the ground state variational wave function 
\begin{equation}
\ket{\psi}=\frac{1}{\mathcal{N}}\prod\limits_{i}(u+v\hat{S}^{+}_i\hat{f}_{i-1,i}\hat{f}_{i,i+1})\ket{\psi}_{\textrm{ref}}, \label{varational}
\end{equation}
where $\ket{\psi}_{\textrm{ref}}=\bigotimes_{i}\ket{\downarrow}_i\bigotimes\ket{\text{vac}}$, and $\ket{\text{vac}}$ denotes full occupation of all auxiliary fermions. $u$ and $v$ are variational parameters, and $\mathcal{N}$ is the normalization factor. This wave function at least has the following three advantages: i) It automatically satisfies the Rydberg blockade constraints. ii) It borrows the idea of the BCS wave function such that it can realize a general superposition of all allowed Rydberg configurations. iii) Tuning $v$ can vary the ratio between the Rydberg atom and the ground state atom.  

Moreover, we can define a set of composite spins as $\ket{\Uparrow}_i=\ket{\uparrow}_i\otimes\ket{0}_{i-1,i}\otimes\ket{0}_{i,i+1}$ and $\ket{\Downarrow}_i=\ket{\downarrow}_i\otimes\ket{1}_{i-1,i}\otimes\ket{1}_{i,i+1}$. We note these composite spins are approximations because they do not obey the spin commutation relation due to the binding of fermions. Nevertheless, in terms of these composite spins, the ground state variational wave function is nothing but a spin-polarized state as \cite{pan,Papic2018prb,PXP1}
\begin{equation}
\ket{\psi}=\frac{1}{\mathcal{N}}\prod\limits_{i}(u \ket{\Downarrow}_i+v \ket{\Uparrow}_i).
\end{equation}
This picture provides a helpful physical intuition, especially for the latter construction of many-body scar states. 

We can compute the energy of this variational wave function under the Hamiltonian Eq. \ref{H-gauge}. Without loss of generality, we can set $u=1$, and $v$ becomes the only variational parameter. In a one-dimensional array, by minimizing this single variational parameter, we find $v=-0.636$ at $\Delta=0$, and the overlap between this optimal variational wave function and the ground state obtained by an exact diagonalization calculation on $\sim 18$ sites is as high as $0.99$ \cite{pan}. When $\Delta$ becomes large and positive, it favors a maximum number of Rydberg atoms. However, because of the blockade effects, the number of Rydberg atoms at most is half the total number of sites. Therefore, these Rydberg atoms either sit at even sites or odd sites, giving rise to translational symmetry breaking and two-fold degeneracy\cite{Subir2004prb,Chepiga2019prl,Xie2021prb,Xie2022prb}. Hence, the one-dimensional array exhibits an Ising quantum phase transition as $\Delta$ increases. The Ising transition has been observed experimentally \cite{Browaeys2016nature,Lukin2017nature}, and the critical exponent has also been determined by studying the Kibble-Zurek mechanism through the transition \cite{Lukin2019nature}. With the wave function ansatz Eq. \ref{varational}, the optimization yields $v\gg 1$, and the wave function produces an equal weight superposition of two symmetry-breaking states. Wave functions similar to Eq. \ref{varational} can also capture the ground state of the one-dimensional array with a larger blockade radius \cite{Cheng-Li,pan} and two-dimensional square and triangular lattices.

Here, we would like to pay special attention to atoms in the two-dimensional frustrated lattices, such as the bond Kagome lattice, where spin liquid ground state has been found \cite{experiment_spin_liquid,Ashvin2021prx,Subir2021pnas,Lukin2022prl,Tarabunga2022prl,Giudice2022prb,Ashvin2022prx,Subir2023prb,Ohler2023prr,Cheng-NJP,Tarabunga2023prb,Sun2023prxq,Meng2023prl,Bauer2023pra,Ashvin2023quantum}. The lattice structure is shown in Fig. \ref{gauge} (c), where atoms are placed in bonds of the Kagome lattice (also called Ruby lattice), and the dashed circle denotes the blockade radius. That is to say, among every four links sharing one vertex, only one atom can be excited to the Rydberg state. Under these conditions, there are an extensive number of closely packed configurations containing the maximum number of Rydberg states and satisfying the blockade condition. When Rydberg excitations are favored at large positive detuning, the ground state should be a quantum superposition of these closely packed configurations, as shown in Fig. \ref{spin-liquid}(a). Such a state is a spin liquid state \cite{experiment_spin_liquid,Ashvin2021prx,Subir2021pnas}. 

\begin{figure}[t]
    \centering
    \includegraphics[width=0.48\textwidth]{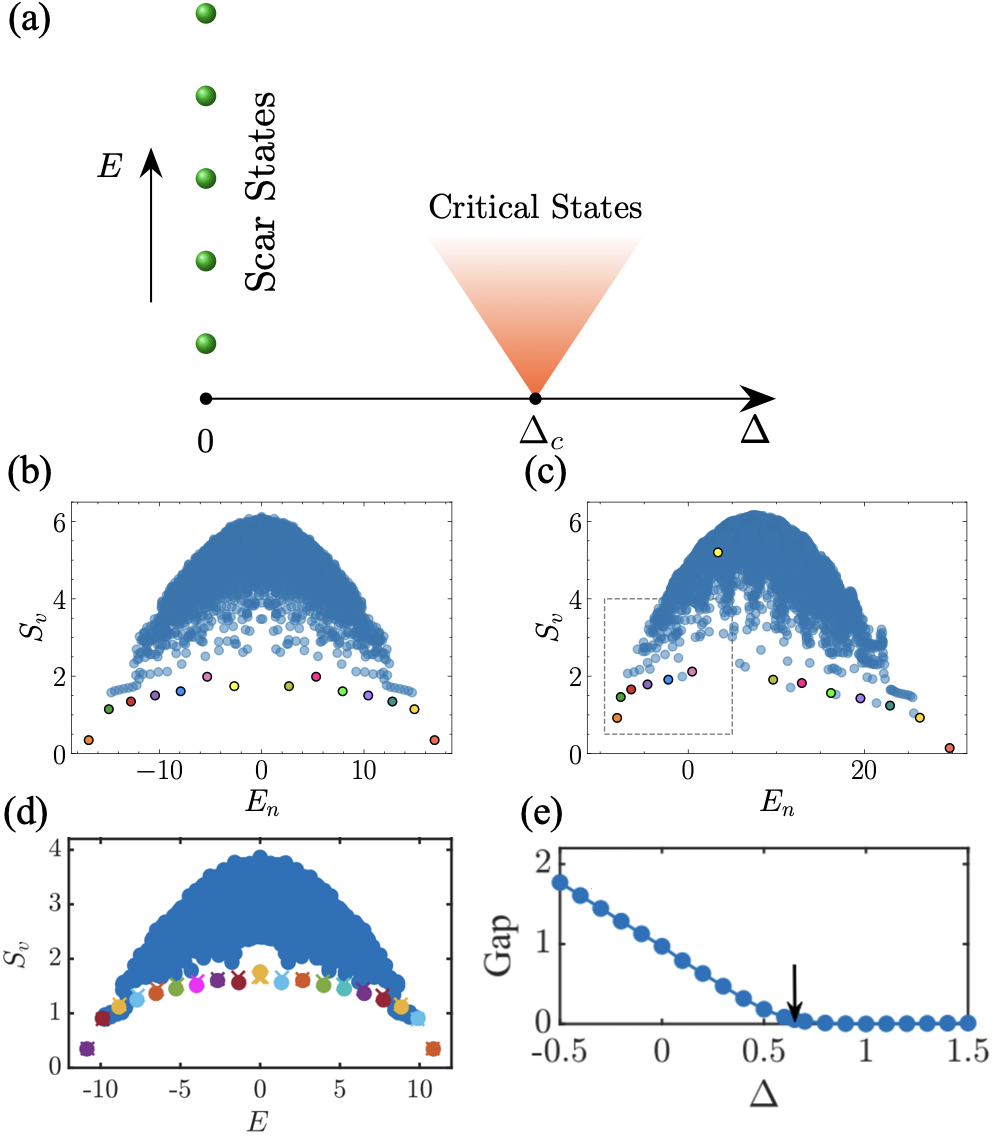}
    \caption{\textbf{Quantum many-body scar states in the PXP model.} (a) Quantum many-body scar states at $\Delta=0$ are non-thermal excited states, and quantum critical states are low-energy states at $\Delta_\text{c}$. Both two states display the sub-volume law entanglement. (b-c) The half-system entanglement entropy for all eigenstates for $\Delta=0$ (b) and $\Delta=\Delta_\text{c}$ (c). In (b), these states marked by different colors are many-body scar states, approximately equally spaced in energy space. Half of these states move toward low energy, as indicated by the dashed box.  (d) At $\Delta=0$, comparison between the half-system entanglement entropy obtained by exact diagonalization (blue points) and that obtained by the magnon variational state (colorful points).  (e) The magnon gap closes at the quantum critical point. (a-c) are reprinted from Ref. \cite{Yao_PRB} and (d-e) are reprinted from Ref. \cite{pan}.
    \label{scar}}
\end{figure}

Following the similar discussion in the above section, the Rydberg blockade condition in the Kagome lattice can also be implemented by introducing auxiliary fermions defined at all sites of the Kagome lattice, as shown in Fig. \ref{gauge}(c) and (d). The gauge theory model is now written as \cite{Cheng-NJP} 
\begin{equation}
\hat{H}_{\text{gauge}}=\frac{\Omega}{2}\sum\limits_{ij}(\hat{S}^{+}_{\langle ij\rangle}\hat{f}_{i}\hat{f}_{j}+\text{h.c.})-\Delta\sum_{\langle ij\rangle}\hat{S}^z_{\langle ij\rangle}. \label{H-gauge-2}
\end{equation} 
Here, for convenience, we use $i$ and $j$ to denote the site index and $\langle ij\rangle$ denotes the link between $i$-site and $j$-site. As shown in Fig. \ref{gauge} (c) and (d), when an atom at the link $\langle ij\rangle$ is excited from the ground state ($\ket{\downarrow}$) to the Rydberg state ($\ket{\uparrow}$), fermions at both $i$-site and $j$-site become occupied, which forbids Rydberg excitation in other six links sharing the site-$i$ or site-$j$. Similar as Eq. \ref{varational}, we can write down a variational wave function as
\begin{equation}
\ket{\psi}=\frac{1}{\mathcal{N}}\prod\limits_{\langle ij\rangle}(u+v\hat{S}^{+}_{\langle ij\rangle}\hat{f}_{i}\hat{f}_{j})\ket{\psi}_{\textrm{ref}}. \label{varational-1}
\end{equation}
This wave function naturally describes a superposition of all configurations allowed by the blockade condition. Similarly, we can set $u=1$ and $v$ is the only variational parameter in this wave function.

In solid-state materials, experimental evidence of spin liquids is usually indirect \cite{review_spin_liquid_leon,wen2007,Zhou_review_2017}. In the Rydberg atom array realization of the spin liquid state, a measurement protocol has been proposed to directly measure the long-range coherence due to the superposition of different spin configurations \cite{Ashvin2021prx}. This experiment measured the expectation values of the $\hat{Z}$ operator and the $\hat{X}$ operator along an open and a closed loop \cite{experiment_spin_liquid}. The $\hat{Z}$ operator measures the parity of Rydberg excitations along the loop, and the $\hat{X}$ operator measures the coherence of different spin configurations along the loop, as shown in Fig. \ref{spin-liquid}(b) and (c) \cite{experiment_spin_liquid,Ashvin2021prx}. The difference between the expectation of these two operators along a closed loop and the square of their expectations along an open loop (half of the closed loop) reveal the non-local topological fluctuations, as shown in Fig. \ref{spin-liquid}(e) and (f). This provides much direct evidence of spin liquid. In Fig. \ref{spin-liquid}(g) and (h), we compare the experimental results with the prediction from our variational wave function Eq. \ref{varational-1} where reasonable agreement has been achieved in a broad parameter range. Here, we determine the value of $v$ by fitting the experimental measured Rydberg population, and there is no other fitting parameter used in this comparison. This demonstrates that the variational wave function Eq. \ref{varational-1} inspired by this gauge theory is also a good description of the spin liquid ground state in the bond Kagome lattice. 

\textit{Thermalization and Many-body Scar States.} The experiment reported in Ref. \cite{Lukin2017nature} found that if one starts with an antiferromagnetic initial state $\ket{\uparrow\downarrow\uparrow\downarrow\dots}$, the subsequent time evolution exhibits coherent oscillations when $\Delta=0$. Later, it was pointed out that a set of eigenstates of this Hamiltonian do not obey the eigenstate thermalization hypothesis (ETH), and the number of such eigenstates is proportional to the number of total system sites \cite{Papic2018np}. These particular eigenstates are now called quantum many-body scar states (short noted as scar states below) as a novel phenomenon of partial violation of quantum thermalization. The antiferromagnetic state has a large overlap with some of these scar states, therefore, the violation of thermalization manifests itself in the time dynamics following the antiferromagnetic initial state. Recently, extensive studies have been made on scar states in the PXP model, AKLT-like and the Hubbard-like models \cite{Bernevig2017sci,Lukin2017nature,Papic2018np,Papic2018prb,papic2018prb1,AKLT1_PRB2018,AKLT2_PRB2018,Papic2019prl,PXP_condensation2019,Lukin2019prl,Motrunich2019PRL,Thomas2019PRL,AKLT2020prr,PXP1,Papic2020prb,AKLT_PRB2020,eta_pairing_scar,Bernevig2020prb,Bernevig2020prb2,Papic2021np,Papic2021prx,Bernevig2022}.  

When an eigenstate obeys ETH, its entanglement entropy should equal the thermal entropy and obey the volume law.  The scar states violate ETH, and their entanglement entropy displays the sub-volume law behavior logarithmically depending on the sub-system size in one dimension \cite{Bernevig2017sci,Papic2018np,Papic2018prb,Motrunich2019PRL,AKLT2_PRB2018,Thomas2019PRL,AKLT2020prr}. On the other hand, we have discussed above that the one-dimensional Rydberg atom array displays an Ising quantum phase transition, and the quantum critical states also display the sub-volume law entanglement entropy \cite{Kitaev2003prl,Cardy2004}. Therefore, it naturally raises the question of whether there exists any relation between the scar states at $\Delta=0$ and the critical states at $\Delta_\text{c}$, as shown in Fig. \ref{scar}(a). However, the former are generically high-energy excited states, and the latter are low-energy states.

\begin{figure}[t]
    \centering
    \includegraphics[width=0.45\textwidth]{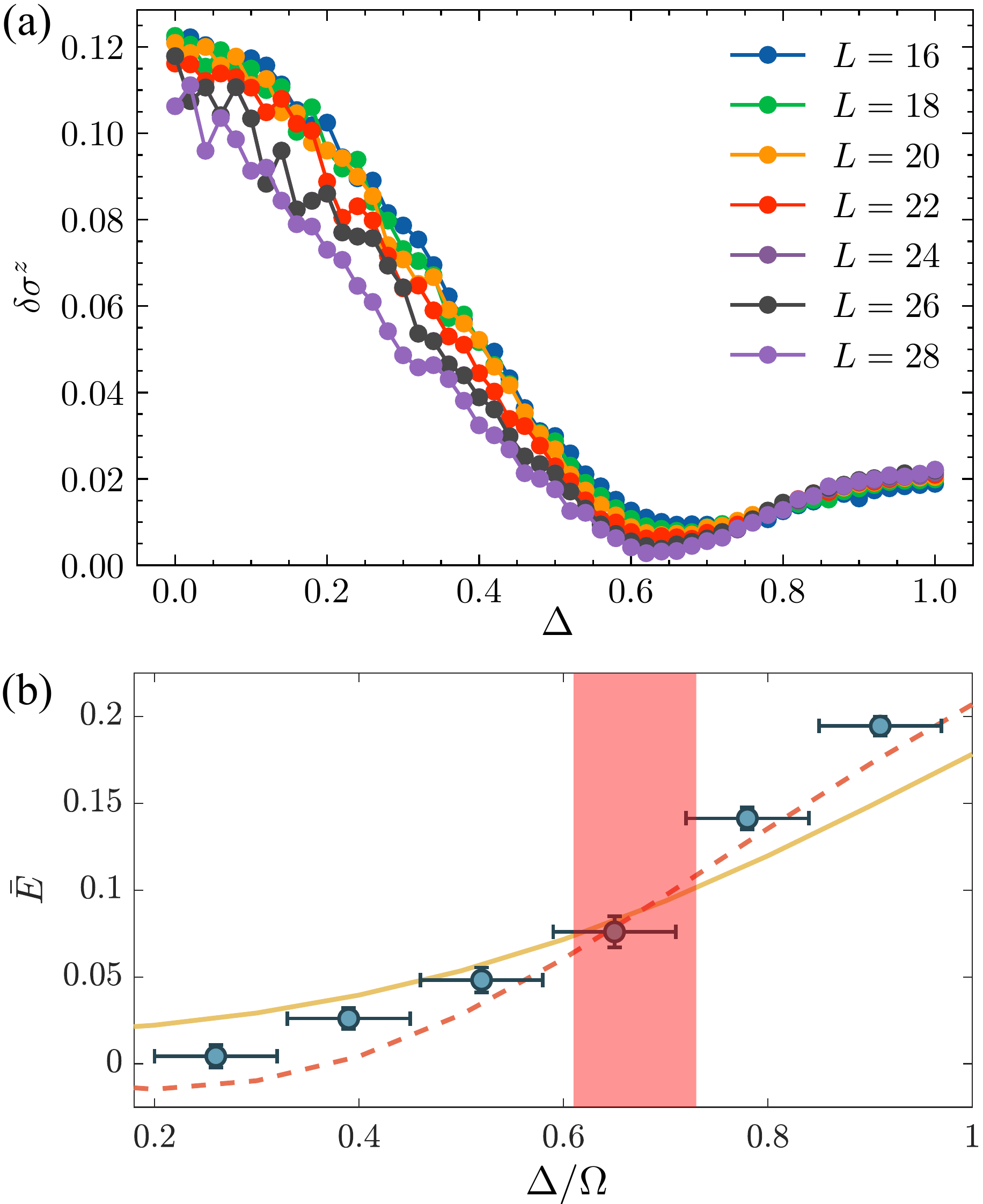}
    \caption{\textbf{Thermalization across the Ising transition}. Starting from the antiferromagnetic initial state, we compare the physical observable $\langle\sigma_z\rangle$ under the long-time steady state and the expected thermal value. (a) is the theoretical result of the difference between these two values and (b) is the experimental measured physical observable under a long-time steady state, compared with the expected thermal value (solid line). (a) is reprinted from Ref. \cite{pan} and (b) is reprinted from Ref. \cite{USTC-thermalization}.
    \label{thermalization}}
\end{figure}

In fact, there exists an adiabatic connection between scar states and critical states  \cite{Yao_PRB,Cui2022PRB,Papic2023PRB}. As $\Delta$ increases toward $\Delta_\text{c}$, a group of scar states move toward the low-energy and become the critical states well described by the Ising conformal field theory, as shown in Fig. \ref{scar}(b-c)  \cite{Yao_PRB}. To understand this result, we recall that the ground state of the PXP model can be well described by the polarized state of the composite spins. Then, we can define a set of $\eta$ operators that flip the composite spin as 
\begin{equation}
\hat{\eta}^\dag_i(u \ket{\Downarrow}_i+v \ket{\Uparrow}_i)=(v\ket{\Downarrow}_i-u \ket{\Uparrow}_i).
\end{equation} 
We can construct the magnon operator with momentum $k$ as $\hat{\eta}^\dag_k=(1/N_\text{s})\sum_i e^{ik R_i}\hat{\eta}^\dag_i$, where $N_\text{s}$ is total number of sites. For single magnon excitation, the minimum excitation energy occurs at $k=\pi$ \cite{pan,PXP_condensation2019}. It can be shown that the multiple magnon states $\{\eta_\pi^\dag\ket{\text{GS}}, (\eta_\pi^\dag)^2\ket{\text{GS}}, (\eta_\pi^\dag)^3\ket{\text{GS}},\dots\}$ have significantly large overlap ($\gtrsim 0.9$) with the actual scars states obtained by exact diagonalization and their entanglement entropy also agree very well, as shown in Fig. \ref{scar}(d) \cite{pan}. This construction of scar states in the PXP model is similar to the $\eta$-operator construction \cite{yang1989PRL,yang1990,zhang1990PRL} of scar states in the Hubbard-like models \cite{eta_pairing_scar,Bernevig2020prb,Bernevig2020prb2} and naturally explains why these scar states are approximately equally spaced in the energy space. It is therefore not surprising to see that the magnon excitation gap closes at the Ising transition, as shown in Fig. \ref{scar}(e), and the magnon carries momentum $k=\pi$ that is consistent with an antiferromagnetic ground state after the transition. Because the scar states can be approximated by magnon excitations, the magnon gap closing drives the scar states toward quantum criticality. 

This relation between scar states and critical states brings out the connection between thermalization and criticality. If we start with an antiferromagnetic initial state $\ket{\uparrow\downarrow\uparrow\downarrow\dots}$, at $\Delta=0$, the long-time evolution does not lead to full thermalization because of the scar states. By increasing $\Delta$ toward $\Delta_\text{c}$, the scar states merge into the low-energy critical states, and the antiferromagnetic initial state thermalizes. Finally, when $\Delta\gg\Delta_\text{c}$, this initial state fails to thermalize again because it becomes the symmetry-breaking ground state. Hence, the antiferromagnetic initial state thermalizes only at the critical regime \cite{Yao_PRB}, and this prediction can be verified by comparing the saturation value of a physical observation after long-time evolution and its expected thermal value, as shown in Fig. \ref{thermalization}(a). This prediction has been experimentally confirmed in the lattice Schwinger model realized in a one-dimensional optical lattice, as shown in Fig. \ref{thermalization}(b) \cite{USTC-thermalization}.

\begin{figure}[t]
    \centering
    \includegraphics[width=0.48\textwidth]{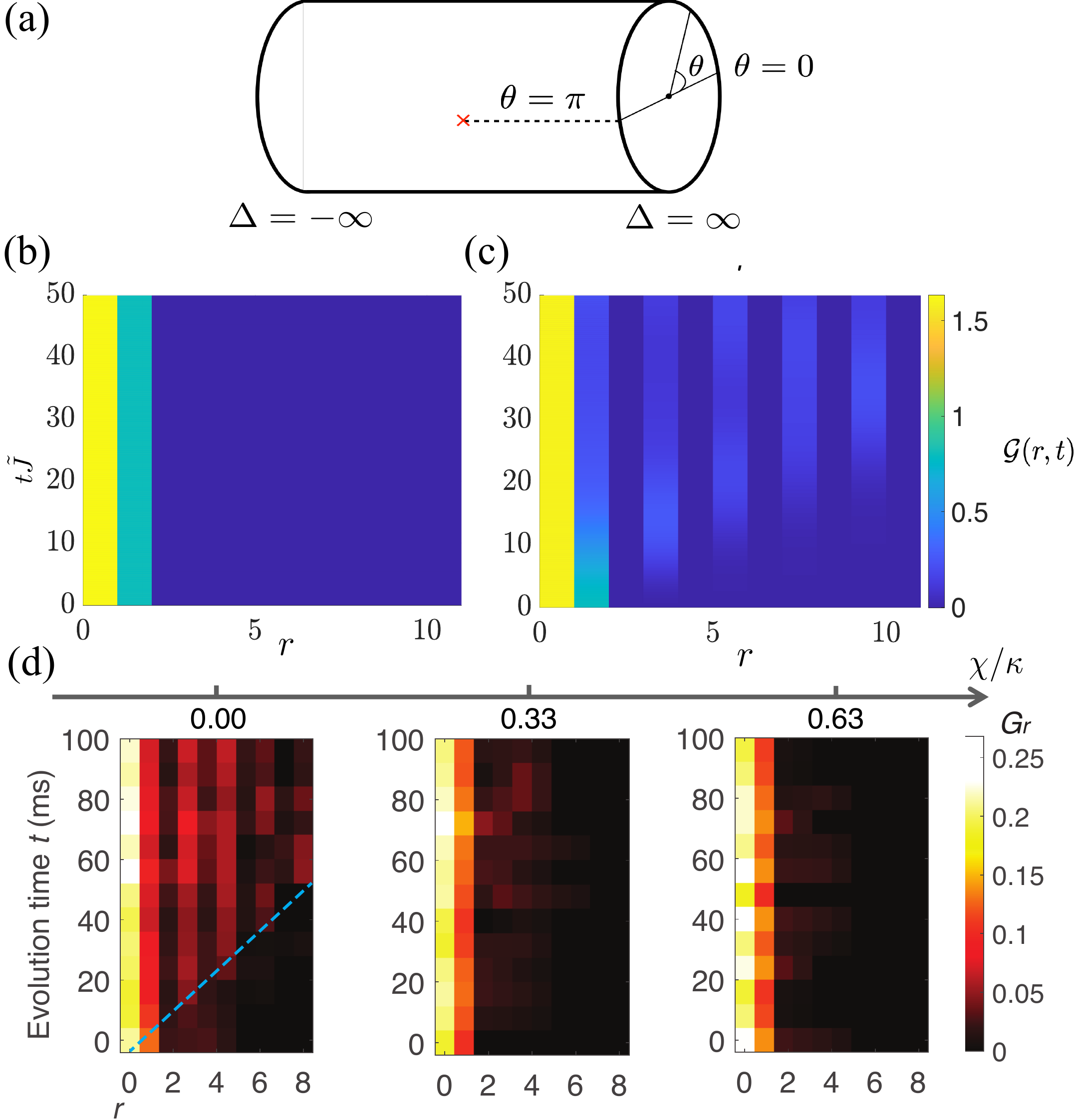}
    \caption{\textbf{Confinement-Deconfinement Dynamics across Ising transition}. (a) The phase diagram of the one-dimensional Schwinger model, where $\theta$ is the topological angle. (b-c) The spin-spin correlation function to detect confinement for $\Delta<\Delta_\text{c}$ (b) and deconfinement for $\Delta>\Delta_\text{c}$ and $\theta=\pi$ (c). (d) Experimental measurement of the spin-spin correlation function in the PXP model. Confinement occurs when the topological angle is tuned away from $\theta=\pi$. (a-c) is reprinted from Ref. \cite{Cheng-PRXQ} and (d) is reprinted from Ref. \cite{USTC-confinement}.
    \label{confinment}}
\end{figure}

\textit{Dynamics and Confinement.} As we have discussed above, the one-dimensional PXP model can be mapped into the lattice Schwinger model. The phase diagram of the lattice Schwinger model is shown in Fig. \ref{confinment}(a) \cite{David}. There are two notable features in this phase diagram. First, there is a parameter $\theta$ called the topological angle. Secondly, when $\theta=\pi$, there is a confinement-deconfinement transition, and the system enters a deconfinement phase when $\Delta>\Delta_\text{c}$. If $\theta\neq\pi$, there is no transition and the quantum phase is always a confined phase. It will be interesting to discuss how these two features manifest in the physical system of Rydberg atom arrays. These two features have also been discussed in the quantum simulation of the lattice gauge theory \cite{Zohar2011prl,Peter2012prl,Peter2017njp,PXP-gauge,Cheng-PRXQ,Hauke}. 

To this end, we take a close look at the conservation law Eq. \ref{conserved-quantity}. We introduce the electric field as $\hat{E}_{i}=(-1)^{i}\hat{S}_{i}^z$, the physical charge as $\hat{\rho}^m_{i,i+1}=(-1)^i\hat{n}^f_{i,i+1}$ and the gauge charge (or the background charge) as $\hat{\rho}^g_{i,i+1}=0$, the Eq. \ref{conserved-quantity} can be mapped to a discrete version of the Gauss's law 
\begin{equation}
\hat{E}_{i+1}-\hat{E}_{i}=\hat{\rho}^m_{i,i+1}+\hat{\rho}^g_{i,i+1}. \label{Gaussian}
\end{equation}
This mapping to Gauss's law can help us to resolve the issue proposed above. First of all, we note that there should be a Maxwell term $g\sum E^2_i$ in the gauge theory. When $g$ is large, the Maxwell term forces $|E_i|$ to take the smallest value at low energy. Note that the topological angle is defined through $E_{i}=n-\frac{\theta}{2\pi}, n\in {\bf Z}$. When $\theta=\pi$, $E_i$ takes $\pm 1/2$ at the low-energy space. This is consistent with our definition $\hat{E}_{i}=(-1)^{i}\hat{S}_{i}^z$. Under this circumstance, the Maxwell term is a constant, and therefore, does not need to be included explicitly. 

To change the topological angle $\theta$, we can redefine $\hat{E}_{i}=(-1)^{i}\hat{S}_{i}^z+h$ with $h\subseteq [0,1]$ being site independent constant \cite{Cheng-PRXQ,Hauke}. With this definition, Eq. \ref{Gaussian} is still consistent with Eq. \ref{conserved-quantity}. However, the accessible values of $E_i$ vary as $h$ changes, and therefore, the topological angle can be tuned by changing $h$. When $h\neq 0$, the Maxwell term contains a staggered magnetic field term $\delta \sum_i(-1)^i\hat{S}_i^z$ ($\delta=2gh$) that cannot be ignored. In other words, if we add this staggered magnetic field into the PXP model, the corresponding gauge theory should acquire a topological term with $\theta\neq \pi$, and the strength of this staggered magnetic field can tune the derivation of the topological angle $\theta$ from $\theta=\pi$ \cite{Cheng-PRXQ,Hauke}.

We note in the phase diagram shown in Fig. \ref{confinment}(a), the confinement-deconfinement transition only occurs at $\theta=\pi$. This feature is also consistent with the Ising transition we discussed above. Because the Ising symmetry in the PXP model is manifested as translational symmetry of one lattice space, this symmetry is explicitly broken by adding the staggered magnetic field term, and therefore, the Ising transition also disappears once the staggered magnetic field.

As described by the Hamiltonian Eq. \ref{H-gauge}, when a spin is suddenly flipped, it simultaneously creates (or annihilates) two fermions at its neighboring links. Because of the definition $\hat{\rho}^m_{i,i+1}=(-1)^i\hat{n}_{i,i+1}^f$, these two fermions have opposite physical charges, and therefore, they can be viewed as a pair of an electron and a positron. Then, confinement means that the electron and the positron are always bound together during the subsequent time evolution, and deconfinement means that the electron and the positron are free to move separately \cite{Cheng-PRXQ}. This effect can be detected by measuring the time evolution of the spatial correlation function of these two physical charges after the sudden spin flip. However, since the fermions in the Hamiltonian are auxiliary, their correlation cannot be directly measured. Nevertheless, thanks to the local conservation law Eq. \ref{conserved-quantity}, we can write the correlation function of physical charge equivalently to the correlation function of spins, which can reveal the confinement-deconfinement transition, as shown in Fig. \ref{confinment}(b) and (c) \cite{Cheng-PRXQ}. This effect can also be extended to one-dimensional Rydberg atom arrays with longer blockade ranges \cite{Cheng-Li}. This confinement-deconfinement dynamics, as well as the proposal of tuning topological angle, has been observed in an optical lattice-simulated PXP model \cite{USTC-confinement}. Some typical experimental results are shown in Fig. \ref{confinment}(d), agreeing with the prediction made in Ref. \cite{Cheng-PRXQ}.

\textit{Outlook.} The strong correlation effects in the Rydberg atom arrays are rooted in the Rydberg blockade effect, which is the origin of rich equilibrium and dynamic phenomena in this platform. This review summarizes a lattice gauge theory implementation of the Rydberg blockade effect and discusses its applications. 

First, Rydberg atom arrays provide a new direction to study strongly correlated many-body physics. Here, motivated by this gauge theory description, we show that a class of variational wave functions can capture the ground state properties of different array geometries, including the spin liquid state in the Kagome lattice. An important further direction is to realize quantum states with various topological orders in this platform, which is closely related to neutral-atom array-based quantum computation. 

Secondly, the quantum many-body scar state was first discovered experimentally in the Rydberg atom arrays. This is an important finding for quantum thermalization. However, many phenomena about scar states are still unclear, for instance, the response of the scar states to Floquet driving \cite{scar-Floquet} and generalizing the scar states to higher dimensions. Later, the scar states were also found in several other strongly correlated models. However, a universal and deep understanding of scar states across different models is still lacking. Here, we show that the gauge theory description can provide a simple physical picture of scar states in this system, helping sharpen our understanding of scar states. 

Thirdly, the Rydberg atom arrays also exhibit rich quantum dynamics. Here, we show an example that the confinement-deconfinement physics of gauge theory can be probed by studying quench dynamics. Other interesting dynamics can include periodically driving the Rydberg excitations and the dissipation dynamics due to the spontaneous decay of Rydberg states. 

Finally, the most exciting development in this platform comes from its great potential for fault-tolerant universal quantum computation\cite{Lukin2022nature,Lukin2023nature,Scholl2023,Ma2023,Lukin2023nature2}. Recent work has also revealed the connection between local gauge symmetry and quantum error correction code \cite{Nathan2023NPJ}. In this platform, both many-body correlation and two-qubit CNOT gate originate from the same Rydberg blockade effect. Hence, we hope that understanding emergent gauge symmetry from correlation effects can eventually help us with quantum error correction in this platform. 

\textit{Acknowledge.} This work is supported by the Innovation Program for Quantum Science and Technology 2021ZD0302005, the Beijing Outstanding Young Scholar Program, the XPLORER Prize, NSFC Grant No. 12204034, No. 12374251 and  Fundamental
Research Funds for the Central Universities (No.FRFTP-22-101A1).

\bibliography{bibli}

\end{document}